\documentclass[conference]{IEEEtran}
\IEEEoverridecommandlockouts
\usepackage{cite}
\usepackage{amsmath,amssymb,amsfonts}
\usepackage{algorithmic}
\usepackage{graphicx}
\usepackage{textcomp}
\usepackage{xcolor}
\usepackage{todonotes}
\usepackage{float}

\usepackage{array}

\def\BibTeX{{\rm B\kern-.05em{\sc i\kern-.025em b}\kern-.08em
    T\kern-.1667em\lower.7ex\hbox{E}\kern-.125emX}}
\begin{document}

\title{Multimodal Detection of Information Disorder \\from Social Media

%
}

\author{\IEEEauthorblockN{Kirchknopf Armin}
\IEEEauthorblockA{\textit{Institute of Creative Media Technologies} \\
\textit{University of Applied Sciences}\\
\textit{St. Pölten}, Austria \\
armin.kirchknopf@fhstp.ac.at}
\and
\IEEEauthorblockN{Slijepčević Djordje}
\IEEEauthorblockA{\textit{Institute of Creative Media Technologies} \\
\textit{University of Applied Sciences}\\
\textit{St. Pölten}, Austria  \\
djordje.slijepcevic@fhstp.ac.at}
\and
\IEEEauthorblockN{Zeppelzauer Matthias}
\IEEEauthorblockA{\textit{Institute of Creative Media Technologies} \\
\textit{University of Applied Sciences}\\
\textit{St. Pölten}, Austria  \\
matthias.zeppelzauer@fhstp.ac.at}
}

\maketitle

\begin{abstract}

Social media is accompanied by an increasing proportion of content that provides fake information or misleading content, known as information disorder. 
%
In this paper, we study the problem of multimodal fake news detection on a large-scale multimodal dataset. We propose a multimodal network architecture that enables different levels and types of information fusion. In addition to the textual and visual content of a posting, we further leverage secondary information, i.e. user comments and  metadata. We fuse information at multiple levels to account for the specific intrinsic structure of the modalities. Our results show that multimodal analysis is highly effective for the task and all modalities contribute positively when fused properly.

\end{abstract}

\begin{IEEEkeywords}
Social Media Retrieval, Multimodal Modeling, Information Disorder, Fake News Detection
\end{IEEEkeywords}

\section{Introduction}

During events like the U.S. presidential election in 2016 the public has become aware of the impact that \textit{fake news} have on public opinion. The identification of such information is highly complex, semantically demanding, and even for experts a difficult task. Due to the ever-increasing amount of data, automated analysis approaches are necessary to assist the detection and verification of fake news.


In the context of this paper\footnote{Funded by netidee, the funding program of the Internet Stiftung.}, we focus on fake news in terms of \textit{information disorder} as defined by Wardle~\cite{wardle_fake_2017}. Three types of information disorder can be distinguished: (i) misinformation, which refers to misleading content produced without a specific intent (ii) disinformation, which refers to purposely generated and potentially harmful content, i.e. false or manipulated content and (iii) malinformation, which is harmful content including hate speech and harassment. 


 \begin{figure}[t]
  \centering
  \includegraphics[width=0.6\columnwidth]{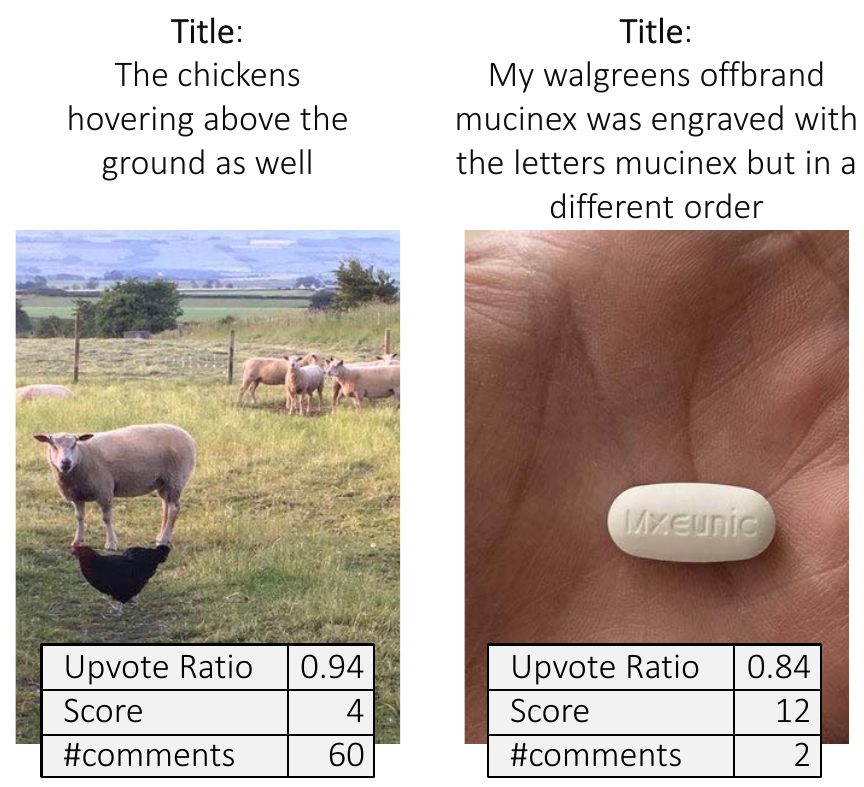}
  \caption{Two samples from our study. The left image shows an example of intended disinformation. 
  The shadow below the sheep in the foreground is a retouched version of a chicken. In the right image, the title states correctly that there is a spelling mistake in the word ``Mucinex'', which is confirmed by the picture (non-fake). These examples show that information disorder has a strong semantic dimension making it challenging to assess even for humans.
  }
  \label{fig:samples}
  \vspace{-10pt}
\end{figure}

Our contribution is an end-to-end learnable modular approach, which combines multiple heterogeneous modalities for the detection of information disorder. To this end, we propose a multimodal multistream network architecture that learns from four heterogeneous input modalities, i.e. two textual and a visual modality, as well as metadata information (Figure~\ref{fig:samples} shows data samples used in our study). We propose to fuse these four structurally different modalities at multiple levels to optimally account for the information contained in each modality. We investigate which modality is most important for the detection of information disorder and whether a combined multimodal analysis is beneficial in contrast to monomodal processing. Our evaluation, conducted on a large-scale multimodal real-world dataset from Reddit~\cite{nakamura_rfakeddit_2020}, shows that multimodal processing strongly improves detection results. This leads us to two conclusions: (i)~all modalities can provide useful clues for the detection of fake news and (ii)~the proposed multilevel hierarchical information fusion allows to successfully capture information from all modalities.








\section{Related Work}

Table~\ref{tab:overview_author_modalities} lists different approaches developed for information disorder detection, including approaches for mis- and disinformation detection, rumor verification, and fake news detection. 
The related literature can be split into two groups, monomodal approaches \cite{ ma_rumor_2018, mohtarami_automatic_2018, lago_visual_2019} and multimodal approaches \cite{singhal_spotfake:_2019, wang_eann:_2018, nakamura_rfakeddit_2020, dong_dual_2018, zubiaga_exploiting_2017, ruchansky_csi:_2017, cui_same:_2019, jin_multimodal_2017}.

\begin{table}[t]
\centering
\caption{Overview of recent related work on fake news detection using either one or a combination of multiple modalities.}
\label{tab:overview_author_modalities}
\begin{tabular}{l|c|c|c}

\multicolumn{1}{c|}{\textbf{Author}}         & \begin{tabular}[c]{@{}c@{}}Textual\\Content\end{tabular} & \begin{tabular}[c]{@{}c@{}}Visual\\Content\end{tabular} &  \begin{tabular}[c]{@{}c@{}}Metadata\end{tabular}  \\ \hline
Ma et al. (2018)  \cite{ma_rumor_2018}           & x    &        &      \\ 
Mohtarami et al. (2018)  \cite{mohtarami_automatic_2018}            & x    &        &      \\ 
Lago et al. (2019)   \cite{lago_visual_2019}                &      & x      &      \\ \hline
Ruchansky et al. (2017) \cite{ruchansky_csi:_2017}             & x    &        & x    \\ 
Zubiaga et al. (2017) \cite{zubiaga_exploiting_2017}& x    &        & x    \\ 
Dong et al. (2018)  \cite{dong_dual_2018}                 & x    &        & x    \\ 
Wang et al. (2018)  \cite{wang_eann:_2018}              & x    & x      &      \\ 
Singhal et al. (2019)  \cite{singhal_spotfake:_2019}              & x    & x      &      \\ 

Nakamura et al. (2020)  \cite{nakamura_rfakeddit_2020}             & x    & x      &      \\ 
Jin et al. (2017) \cite{jin_multimodal_2017}                   & x    & x      & x    \\ 
Cui et al. (2019) \cite{cui_same:_2019}                   & x    & x      & x    \\ 
Papadopoulou et al. (2019) \cite{papadopoulou2019corpus}                   & x    & x      & x    \\ 
\end{tabular}
\vspace{-10pt}
\end{table}

Ma et al.~\cite{ma_rumor_2018} propose a text-based method for rumor detection by combining propagation trees with RNNs. The text is thereby modeled as a time-series. Mohtarami et al.~\cite{mohtarami_automatic_2018} propose a text-based method based on similarity modeling and stance filtering, which can extract text portions that can explain the factuality of a given claim. Lago et al.~\cite{lago_visual_2019} propose and evaluated different methods for identifying manipulated images in a dataset by using image forensic methods, moving the focus from textual to visual data. 

Fake information can span multiple modalities, thus, multimodal approaches represent our main focus. Singhal et al.~\cite{singhal_spotfake:_2019} propose a multimodal framework for fake news detection that employs a language transformer and visual models (pretrained CNNs). The modality-specific feature representations are fused by concatenation and fed into a binary classifier. Wang et al.~\cite{wang_eann:_2018} 
propose an event independent fake news detector based on an adversarial network. The approach utilizes a two-stream feature extractor, one for text, based on a text-based CNN, and a VGG-19 model for the visual modality. Both representations are fused by concatenation. Similarly, Nakamura et al.~\cite{nakamura_rfakeddit_2020} utilize a two-stream network for processing textual and visual information. A bidirectional BERT encodes text data and a ResNet50 model is used for visual data. The resulting embeddings are fused by taking their element-wise maximum.

Contrary to previously mentioned approaches, Dong et al.~\cite{dong_dual_2018} utilize textual information and user-based metadata, such as age and number of followers. The features, are combined and processed by and attention-based bidirectional Gated Recurrent Unit network. The extracted features are then fed into an unified attention model  and patterns in the attention distribution are leveraged to detect fake news. Similarly, Zubiaga et al.~\cite{zubiaga_exploiting_2017} combine text content and metadata for rumor detection. For this purpose the authors utilize Conditional Random Fields, which enable to model neighboring information (context). Ruchansky et al.~\cite{ruchansky_csi:_2017} introduce a three step approach consisting out of (i) fetching and modelling user interactions together with the related text, (ii) creating a score which determines how suspicious individual users are and (iii) combining the previous steps, the user features, the text features and the score, for predicting a label. Modelity-specific features are fused by concatenation.

Only a few approaches leverage textual, visual and metadata. Cui et al.~\cite{cui_same:_2019} propose a  multistream architecture with  an adversarial loss, which individually influences all three network branches. The fused features are fed into a fully connected layer followed by a softmax to obtain likelihoods for fake news. Jin et al.~\cite{jin_multimodal_2017} propose a multimodal approach with a multilevel fusion setup similarly to our approach. First, word embeddings from the text and social context features are concatenated and fed into an LSTM. In parallel, a VGG-19 model extracts visual features, which are then multiplied (element-wise) by the attention output of the LSTM network. The word embeddings and the visual features are then concatenated and processed by a binary classifier. In contrast to this approach, we model metadata in a separate branch and add an additional stream for secondary data (e.g. comments). A multimodal dataset composed of Youtube videos, titles and metadata has been introduced by Papadopoulou et al.~\cite{papadopoulou2019corpus}. 

\section{A Multimodal Approach for  Information Disorder Detection}

Information disorder is a semantically complex concept that manifests itself in different modalities. We assume that the fusion of information from multiple modalities is important to solve this task. We propose an approach for information disorder detection based on \textit{four} input modalities, namely (i) primary textual content, i.e. the suspicious posting or news item itself, (ii) secondary information, i.e. content about the primary content (e.g. comments), (iii) the visual content of the posting, and (iv) available metadata about the other modalities. A particular challenge is to fuse the information from these different types of inputs, which differ not only structurally (e.g. text vs. image) but also in dimensionality (e.g. high-dimensional visual embeddings vs. low-dimensional abstract data in case of metadata).

\begin{figure}[t]
  \centering
  \includegraphics[width=1\columnwidth]{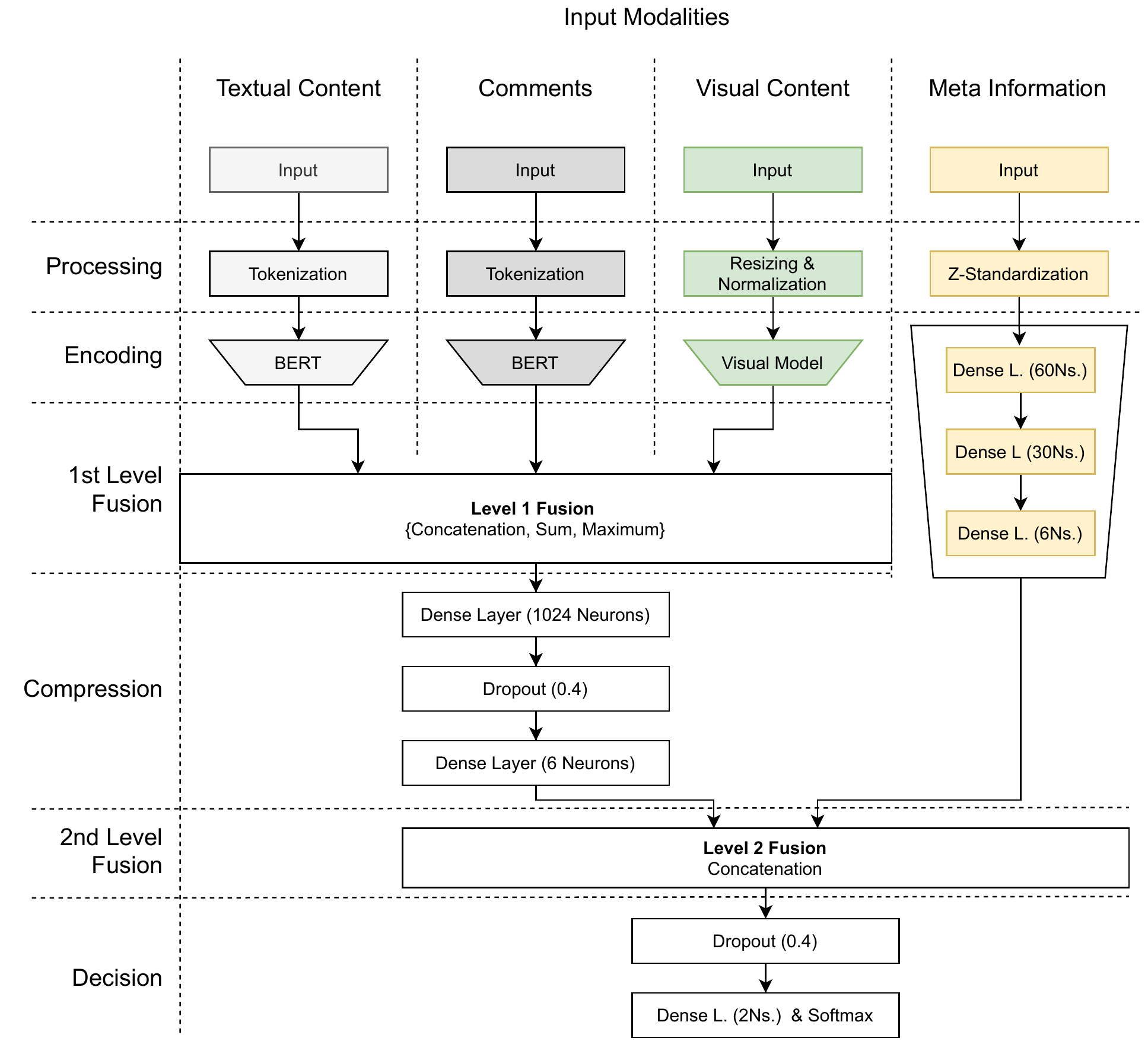}
  \caption{Architectural overview of the proposed multimodal information disorder detection method.}
  \label{fig:overview_method_detailed}
  \vspace{-10pt}
\end{figure}

The proposed multimodal network architecture is depicted in Figure~\ref{fig:overview_method_detailed}. Each modality is processed by one separate branch (stream). The first stream takes the actual content of a social media posting as input (e.g. the title and, if available, its body). The second stream processes textual information \textit{related} to the posting, e.g. the comments available for the post. To keep the representation simple and comparable to the first stream, we concatenate all available comments to obtain one consolidated input. It is important to note that both textual modalities capture different perspectives on the actual content and are modeled in separate branches. 
We use a similar processing chain for both textual modalities, the dataset authors prepared a cleaned version of the textual features, so no preprocessing was necessary.  A BERT model is used to obtain separate text embeddings for the two inputs.  

The third stream processes the visual modality. First the images are standardized to zero-mean by calculating the mean over the entire training set (per channel) and subtracting it. After normalizing them to [0,1], the images are passed to a pre-trained CNN to obtain a feature representation. Theoretically, an arbitrary visual encoder network can be employed. In our case, we investigate three state-of-the-art image classification networks, i.e.~ResNet-v2, ResNet101-v2, and Inception-v3.

The fourth and final stream takes a vector of metadata as input. It may contain social media metrics or categorical data, e.g. the number of comments, the number of likes/dislikes, the number of upvotes or other ranking information. The metadata first need to be normalized to a well-defined value range and then concatenated into a vector. Since no pre-defined encoder for such data exists, we propose to train a light-weight multilayer perceptron (MLP) to represent the input data. We stack three dense  layers with 60, 30, and 6 neurons, respectively and ReLU activation functions.  

The individual processing streams produce representations of different dimension. Thus, we propose a hierarchical scheme to fuse the information of the different modalities (see Figure~\ref{fig:overview_method_detailed}). This prevents that higher-dimensional representations (e.g. from BERT and the visual CNNs) dominate the other (lower-dimensional) representations like the one obtained from the metadata. The first fusion level combines the textual and visual representations. These embedding vectors are designed to have all equal length (and thereby equal relevance in the fusion). This allows the use of different fusion strategies like concatenation, element-wise maximum of input vectors and element-wise average over all input vectors. Since it is not clear, which of these fusion operations is most beneficial, we evaluate them systematically in our experiments.

The fused information is then further compressed by a stack of dense layers, so that it matches the dimensionality of the representation obtained by the fourth stream. At the second fusion level the two remaining representations are concatenated. Thereby, we provide more influence to the metadata modality on the final detection (equal balance of content and metadata), which has shown to be beneficial in practice (other weightings are possible by adapting the dimensionality of the representations, i.e. the layer size). The final decision is made by a densely connected layer with two output neurons indicating fake vs. non-fake information, followed by a softmax layer to obtain normalized probabilities.


\section{Experiments and Results}


\subsection{Dataset}

For our experiments and study on multimodal fake news detection, we utilize the Fakeddit dataset \cite{ nakamura_rfakeddit_2020}, since it represents the largest published multimodal fake news dataset (to the best of our knowledge). A further advantage is that the authors provide a  split of the dataset into train, validation, and test set, which enables directly comparable experiments. The Fakeddit dataset consists out of one million samples from up to six different categories of information disorder and was collected by the pushshift API. Ground truth exists for binary fake/non-fake classification as well as more fine-grained distinction of 3 and 6 classes, respectively. The dataset contains Reddit postings with comments, with many of the postings contain text and images. Additionally, several metadata attributes are available, such as up- and downvotes of postings, the number of comments, up- and downvote score for each comment, and a score for the post itself. Figure~\ref{fig:samples} depicts two data samples from the dataset.

For our experiments we preprocess the data (similarly to~\cite{nakamura_rfakeddit_2020}) by removing samples where not all modalities are available (e.g. text-only postings), which results in 560622 samples for training, 58972 samples for validation and 58954 hold-out samples for testing. As performance measure we employ accuracy to allow for comparability with Nakamura et al.~\cite{nakamura_rfakeddit_2020}. 
\color{black}

\subsection{Experimental Setup}

To adapt our approach to the input data, we first preprocess the different modalities. For the textual modalities, all related comments for a post are collected from a provided file and preprocessed as described in \cite{nakamura_rfakeddit_2020}. The result is fed into the pre-trained BERT model. The sequence length of BERT is pre-allocated by shortening the input sequences to an average length (calculated over the training set) to reduce training time. This is performed separately for both textual modalities. Images are scaled and normalized. To assess the influence of  different image resolutions, we resize the images to 256x256px and 768x768px, respectively. The metadata attributes from the Fakeddit dataset include the up- and downvotes per post, its score and the count of comments. To normalize the large value range of these attributes we z-standardize all metadata features such as the count of comments and the score, except for the  up- and downvotes, which are already normalized between [0,1]. The  attributes are then provided to the three-layered MLP. We initialized the MLP weights with Glorot
initialization and used L1 and L2 regularization. After the fusion layers we used dropout to prevent overfitting. Training of the multistream network is performed end-to-end via the Adam optimizer ($\beta_1 = 0.9$, $\beta_2 = 0.999$ and $\epsilon = 10^{-8}$) and a learning rate of $10^{-4}$. 

In general, the proposed approach is completely end-to-end trainable. This would, however, result in a large computational effort. As an alternative, each modality can also been trained individually. In our experiments, we achieved the best results by pre-training each modality (stream) separately, and then training only the fusion and classification layers on top (for 10 epochs with a batch size of 96).

We evaluate our approach in multiple steps. First, all modalities are evaluated together, to fully exploit all available information in the dataset and to obtain a multimodal performance baseline. This performance baseline is then compared to the multimodal baseline provided by~\cite{nakamura_rfakeddit_2020}. We compare different fusion variants to estimate the best strategy for information fusion. Subsequently, we evaluate all possible combinations of modalities and further evaluate each modality in isolation to investigate the influence and expressiveness of each modality.



\subsection{Results}

\begin{table}[t]
\caption{Results of our method for different combinations of modalities, fusion strategies and a comparison with the previous baseline.}
\label{tab:overview_results}
\resizebox{1\columnwidth}{!}{%
\begin{tabular}{c|c|c|c|c|c|c|c|c}

\multicolumn{1}{c|}{\#} &
\multicolumn{1}{c|}{Approach} & \begin{tabular}[c]{@{}c@{}}Textual\\Content\end{tabular} & \begin{tabular}[c]{@{}c@{}}Textual\\Comments\end{tabular} &
\begin{tabular}[c]{@{}c@{}}Visual\\Content\end{tabular} &
\begin{tabular}[c]{@{}c@{}}Meta-\\data\end{tabular} & 
\begin{tabular}[c]{@{}c@{}}Fusion\\Strategy\end{tabular}  & 
\begin{tabular}[c]{@{}c@{}}Val.\\Acc.\end{tabular} &
\begin{tabular}[c]{@{}c@{}}Test\\Acc.\end{tabular}
\\
\hline
1          &  Our approach            & x                                                    & x                                                       & x      & x    & Sum     & 95.2\% & 95.5\%  \\ 
2          &Our approach            & x                                                    & x                                                       & x      & x    & Concat.  & 95.0\% & 95.2\%  \\
3          &Our approach            & x                                                    & x                                                       & x      & x    & Maximum & 94.9\% & 95.1\%  \\ \hline
4          &Our approach            & x                                                    & x                                                       & x      &      & Concat.      & 94.9\% & 95.0\%  \\ 
5          &Our approach            &                                                      & x                                                       & x      & x    & Concat.      & 91.2\% & 91.3\%  \\ 
6          & Our approach            & x                                                    &                                                         & x      & x    & Concat.      & 92.8\% & 92.8\%  \\ 
7          & Our approach            & x                                                    & x                                                       &        & x    & Concat.      & 94.4\% & 94.5\%  \\ \hline
8          & Our approach            & x                                                    &                                                         & x      &      & Concat.      & 90.8\% & 91.0\%  \\ 
9          & Our approach            & x                                                    & x                                                       &        &      & Concat.      & 85.9\% & 85.7\%  \\ 
10          &Our approach            & x                                                    &                                                         &        & x    & Concat.      & 88.1\% & 88.2\%  \\ 
11          &Our approach            &                                                      & x                                                       &        & x    & Concat.      & 78.2\% & 78.2\%  \\ 
12          &Our approach            &                                                      &                                                         & x      & x    & Concat.      & 81.1\% & 81.6\%  \\ 
13          &Our approach            &                                                      & x                                                       & x      &      & Concat.      & 88.0\% & 88.1\%  \\ \hline
14          &Our approach            & x                                                    &                                                         &        &      & -      & 88.1\% & 88.1\%  \\ 
15          &Our approach            &                                                      & x                                                       &        &      & -      & 86.7\% & 86.5\%  \\ 
16          &Our approach            &                                                      &                                                         & x      &      & -      & 81.0\% & 81.5\%  \\ 
17          &Our approach            &                                                      &                                                         &        & x    & -      & 77.8\% & 77.3\%  \\ \hline
18          & ~\cite{nakamura_rfakeddit_2020}       & x                                                    &                                                         &        &      & -      & 86.5\% & 86.4\% \\
19          & ~\cite{nakamura_rfakeddit_2020}       &                                                      &                                                         & x      &      & -      & 80.4\% & 80.7\%  \\ 
20          & ~\cite{nakamura_rfakeddit_2020}       & x                                                    &                                                         & x      &      & Maximum      & 89.3\% & 89.1\% 
\end{tabular}
}
\vspace{-7pt}
\end{table}

Table~\ref{tab:overview_results} shows the experimental results for our approach with different selections of modalities. Note that from the different visual encoders evaluated, Inception-v3 with an input resolution of 768x768px provided the best results and only these results are presented in column ``Visual Content''. In rows 4-13, only the best results are reported, which were all obtained by concatenation as fusion strategy. In general, for all evaluated approaches and configurations the generalization to the hold-out test set is high. There is no notable drop in performance between validation and test accuracy. 

For individual modalities (rows 14-17), we observe that the most informative modality is the primary textual content, followed by secondary information (i.e. comments), the visual modality, and metadata. Both, the text-only and the image-only configuration outperform the respective configurations of~\cite{nakamura_rfakeddit_2020} (rows 18-19) and, therefore, represent new performance baselines. The combination of multiple modalities is beneficial in all experiments, showing that the task is truly multimodal.

By combining the two content modalities (text and images) Nakamura et al.~\cite{nakamura_rfakeddit_2020} yield a test accuracy of 89.1\%. Our approach using the same modalities yields 91\% (row 8). Note that this is the best result obtained by using just two modalities, which shows that a special role has to be attributed to the content modalities. Our experiments further reveal that the other modalities (not used by \cite{nakamura_rfakeddit_2020}) can also contribute positively. Adding metadata (row 6) yields 92.8\% and adding the comments as a third modality (row 4) pushes performance to approx. 95\%. The fusion of all four modalities (rows 1-3) surpasses even the 95\% mark. This performance improvement comes at the cost of model complexity, which ranges from 2,450 parameters for the metadata modality, to 177M parameters for textual modalities (BERT), to 380M parameters for full multimodal processing. The improvement over the baseline in~\cite{nakamura_rfakeddit_2020} has (at least) two reasons: i) we use two additional modalities that are useful for the task and ii) we fine-tune all input streams (including the BERT model), which alone yields around 2\% performance gain (from 86\% to 88\%). Finally, from the experiments in rows 1-3 we observe that all three fusion strategies yield similarly good results. 


\section{Conclusion}

We proposed a multimodal architecture for the detection of information disorder, which incorporates not only the content of a social media postings but also metadata and secondary content related to the post. The additional modalities improve performance, which indicates that they contribute useful information. Our evaluation clearly shows that multimodal processing is superior to monomodal processing. The fact that each modality in itself can contribute positively further demonstrates that the proposed multilevel fusion strategy adequately combines the structurally different modalities. In follow-up research, we plan to integrate a social network graph connecting postings, comments, and users as additional modality, which has shown to contribute positively to fake news detection in~\cite{yu2020iarnet}.

\bibliographystyle{IEEEtran}
\bibliography{bib}

\end{document}